\documentclass[journal,10 pt, two column]{IEEEtran}
\usepackage{epsfig}
\usepackage{graphicx}
\usepackage{balance}
\usepackage{float}
\usepackage{psfrag}
\usepackage{amsfonts,amsmath,amssymb,mathrsfs,txfonts}

\usepackage{url}
\usepackage{cite}
\usepackage{verbatim}
\usepackage{multirow}
\usepackage{pifont}
\usepackage{algorithm,algpseudocode}
\usepackage{color}
\usepackage{lettrine}
\usepackage{graphicx}
\usepackage{graphicx}
\usepackage{amsmath}
\usepackage{amsmath,amssymb}
\usepackage{fixltx2e}
\usepackage{hyperref}
\usepackage{adjmulticol}
\usepackage{tabularx}
\graphicspath{{JPG/}}

\usepackage{algorithm,algpseudocode}
\begin{document}
\title{Quantum-Based Solutions for Security Enhancement in Open Radio Access Networks }
\author{Dzung Quoc Ngo, Tharmikka Raveendranathan, Tuan Anh Le, Vinod Sharma, Purav Shah, and Huan X. Nguyen
\thanks{D. Q. Ngo, T. Raveendranathan, T. A. Le, P. Shah, and H. X. Nguyen are with the London Digital Twin Research Centre and 5G/6G Research Group, Middlesex University, London, NW4 4BT, U. K. Email: quocdung4852@gmail.com; tharmikkar@eng.pdn.ac.lk; \{t.le;p.shah;h.nguyen\}@mdx.ac.uk. H. X. Nguyen is also with the College of Engineering and Computer Science, VinUniversity, Hanoi, Vietnam.}
\thanks{V. Sharma is with the School of Engineering, Shiv Nadar University, Delhi NCR, Gautam Buddha Nagar-201314, Uttar Pradesh, India. Email: vinod.sharma@snu.edu.in.}
\thanks{This work is supported by the UK International Science Partnerships Fund
(ISPF), which is managed by the Department for Science, Innovation and
Technology, through two funding programmes: British Council International
Research Collaborations programme with Japan (grant ID: 1583-Nguyen-
Japan) and the UKRI EPSRC (grant ID: UKRI554, led by University of East
Anglia) under the UKI-FNI’s pilot project ‘BEAM-RAN’ (UEA ref: R213867). This work is also supported by the British Council under the Going Global Partnerships programme (grant ID: TNE2025-4791).}
}
\maketitle
\begin{abstract}
 Open Radio Access Networks (O-RAN) introduce unprecedented flexibility, interoperability, and intelligence into next-generation wireless systems, but their disaggregated and software-defined architecture also expands the attack surface and creates new security vulnerabilities. Conventional cryptographic mechanisms, while effective against classical threats, may become insufficient in the presence of quantum-enabled adversaries. This article presents a comprehensive perspective on quantum security for O-RAN, examining how quantum-resilient mechanisms can enhance confidentiality, authentication, and trust across the RAN ecosystem. It discusses post-quantum cryptography (PQC), quantum cryptography, quantum authentication, and quantum-enhanced threat detection within a zero-trust architecture based on continuous verification, least privilege, and micro-segmentation. Their integration with the Near-Real-Time (Near-RT) RAN Intelligent Controller, O-Cloud, and open interfaces is analyzed, together with practical deployment considerations, technology maturity, and adoption timelines. Finally, open research directions are outlined toward secure, resilient, and future-proof O-RAN architectures for 6G networks.
\end{abstract}
\section{Introduction}
The evolution of mobile network architectures toward openness and intelligence is redefining the design principles of next-generation wireless systems. Open Radio Access Networks (O-RAN) represent a paradigm shift from vertically integrated, single-vendor deployments to disaggregated, multi-vendor ecosystems, where the base station is decomposed into functional units, i.e., Radio Unit (RU), Distributed Unit (DU), and Centralized Unit (CU), interconnected via standardized open interfaces. This transformation, driven by the O-RAN Alliance, enables unprecedented flexibility, vendor interoperability, and programmability through the RAN Intelligent Controller (RIC), facilitating real-time optimization via AI/ML-driven xApps and rApps~\cite{10495907}.
However, this openness fundamentally alters the security landscape. The disaggregation of network functions, exposure of open interfaces, e.g., O1, E2, A1, and integration of third-party applications significantly expand the attack surface, introducing new vulnerabilities across control, management, and user planes. Consequently, O-RAN must transition toward a Zero Trust Architecture (ZTA), shifting from perimeter-based security to continuous verification, least privilege, and micro-segmentation. Recent studies highlight critical threats in O-RAN, including unauthorized access, signaling storms, man-in-the-middle (MITM) attacks, and adversarial manipulation of AI/ML pipelines~\cite{10495907},~\cite{10364950}. Unlike legacy RAN systems with tightly coupled security perimeters, O-RAN lacks a unified trust boundary, making end-to-end security enforcement inherently more complex and necessitating fundamentally new security paradigms.

Concurrently, the advent of quantum computing introduces a systemic risk to the cryptographic primitives that underpin modern communication security. Public-key schemes such as Rivest--Shamir--Adleman (RSA) and Elliptic Curve Cryptography (ECC), widely deployed for authentication, key exchange, and digital signatures, are vulnerable to quantum adversaries through algorithms such as Shor’s algorithm, which can efficiently solve integer factorization and discrete logarithm problems~\cite{10233127}. This threat is exacerbated by the “harvest now, decrypt later” model, whereby adversaries passively collect encrypted traffic today for future decryption, posing long-term risks to sensitive data and critical infrastructure. Given the extended operational life-cycles of telecommunications systems, this creates an urgent need for quantum-resilient security mechanisms within O-RAN deployments. These mechanisms fall into two complementary categories: Post-Quantum Cryptography (PQC), which uses quantum-resistant mathematical algorithms on classical systems, and quantum communication, which uses quantum physical properties and dedicated hardware.

Beyond posing threats, quantum technologies offer a fundamentally new toolkit for securing next-generation networks. Quantum cryptographic techniques, such as Quantum Key Distribution (QKD), Quantum Digital Signatures (QDS), and Quantum Identity Authentication (QIA), leverage intrinsic properties of quantum mechanics, including superposition, entanglement, and the no-cloning theorem, to provide provable security guarantees against both classical and quantum adversaries. In parallel, Quantum Machine Learning (QML) introduces powerful capabilities for high-dimensional data processing and anomaly detection, enabling scalable and adaptive security solutions for complex, dynamic network environments~\cite{10233127}.
In this paper, we present a comprehensive framework for integrating quantum-safe mechanisms into O-RAN architectures. The framework places standardized PQC and longer-term quantum technologies inside a zero-trust control model rather than treating them as isolated protocols. Specifically, we investigate how continuous verification, least privilege, micro-segmentation, quantum-safe authentication, secure key establishment, and quantum-driven intelligence can address critical security challenges across O-RAN interfaces, components, and control loops. We further propose practical deployment models, including quantum functionalities as xApps within the RIC and dedicated quantum nodes for secure key management and communication~\cite{10495907},~\cite{9684555}.
The contributions of this work are threefold:
\begin{enumerate}
 \item
 We systematically analyze the security vulnerabilities of O-RAN in the context of emerging quantum threats;
 \item
 We map these vulnerabilities to quantum-safe defense mechanisms aligned with Zero Trust principles, including PQC, communication security, authentication, and intelligent threat mitigation;
 \item
 We propose deployable architectural enhancements for embedding quantum security into O-RAN while preserving its openness and scalability.
\end{enumerate}

By bridging quantum information science with open wireless architectures, this work establishes a forward-looking foundation for secure, resilient, and future-proof O-RAN systems in the quantum era.

\section{Security concerns with O-RAN}

Network security ensures authenticity, integrity, confidentiality, and availability across the network. However, O-RAN's openness and decoupled architecture introduce additional vulnerabilities, including multi-vendor insider threats, while lacking the end-to-end security of conventional RAN. The O-RAN Alliance WG11 addresses these issues by identifying security requirements, modeling threats, and proposing mitigation mechanisms for all nodes and interfaces \cite{10807044}, focusing on open interface data and signaling protection, O-RAN element security, and cloud-native deployment protection. This paper categorizes O-RAN threats into three main groups as follows.

\subsection{Authentication and Authorization}

The open interface and softwarized O-RAN architecture create opportunities for attackers to gain unauthorized access to O-RAN elements by exploiting management-interface vulnerabilities. Unauthorized access to a physical layer comprising RF links, optical fiber, and terminals connected to O-RAN components may cause a critical security breach. An unauthorized device on the fronthaul can flood the physical interface with excessive traffic, resulting in a system crash~\cite{10467183}. It can also deny service by deactivating physical interfaces or overwhelming management, synchronization, control, and user-plane traffic.

Authentication and authorization are essential to O-RAN security. Authentication verifies the identity of users, devices, and applications attempting to connect, while authorization determines their permissions and privileges ~\cite{10495907}. O-RAN's disaggregated, multi-vendor components communicate through standardized interfaces, so each connection requires robust authentication and fine-grained authorization.

Improper policy enforcement or an insecure password-reset mechanism can enable spoofing and unauthorized access. In an elevation-of-privilege attack, an adversary bypasses authentication or access control to perform actions with greater permissions. This can affect virtualization layers, O1, E2, A1, the Service Management and Orchestration (SMO), and the RIC. The E2 interface is particularly important because xApps connect to E2 nodes and access sensitive APIs; excessive xApp permissions can expose the entire Near-RT RIC.

\subsection{ Privacy, Integrity, Confidentiality and Availability}

The core pillars of network security, i.e., privacy, integrity, confidentiality, and availability, are all threatened within O-RAN. Privacy governs data protection and owner rights; integrity ensures data accuracy and completeness; confidentiality restricts access to authorized parties; and availability guarantees continuous network access. Man-in-the-Middle (MITM) attacks are the primary threat to all four pillars, intercepting communication across the fronthaul user, management, and synchronization planes.

The management plane (M-plane) coordinates the user plane (U-plane) and synchronization plane (S-plane). An attacker on the M-plane can impersonate the master clock, a technique known as spoofing, to disrupt network-wide synchronization. In a Precision Time Protocol (PTP) network, a malicious device achieves this by broadcasting fake announcements to claim the master clock role, enabling it to delete, tamper with, or delay PTP packets \cite{10467183}, causing inaccurate synchronization and integrity violations. Similarly, fronthaul control plane spoofing allows attackers to steal and modify uplink/downlink messages, breaching integrity and confidentiality.

The Near-RT RIC is vulnerable through xApp-based E2 node subscription attacks \cite{EuroSP2025-ORAN}. Malicious xApps flood E2 nodes with subscription requests, causing legitimate xApps to miss real-time measurements. Since xApps maintain RIC Message Router (RMR) tables for routing, repeated subscriptions cause the RMR table to route messages to the malicious xApp instead of the intended recipient \cite{EuroSP2025-ORAN}, compromising user equipment (UE) identification, tracking, and priority, violating both integrity and confidentiality. Additionally, a compromised E2 node can enable traffic sniffing across the point-to-point E2 interface.

The A1 interface, which manages AI policy for the Non-Real-Time (Non-RT) RIC, requires bidirectional authentication between the Non-RT RIC and Near-RT RIC. A hostile Near-RT RIC can exploit this to conduct MITM attacks, corrupting Non-RT RIC policies and potentially causing denial of service (DoS).

Availability is primarily threatened by DoS attacks. Synchronization failure via S-plane interception can collapse the entire network \cite{10467183}. Signaling storms can overload and crash the system \cite{10364950}, triggered by flooding initiation/de-registration requests to the core network, high-speed group handovers, remote-reboot malware coordinating mass device restarts, or chatty mobile applications consuming excessive radio resources.

\subsection{Poisoning Attacks-Model and Learning Inaccuracy}
AI/ML used in the Near-RT and Non-RT RIC is vulnerable to data poisoning, extraction, injection, xApp/rApp model theft, logic corruption, and backdoor attacks~\cite{10467183}. In a poisoning attack, an adversary modifies or mislabels training data so that the resulting model makes incorrect decisions. A poisoned model deployed through the SMO can cause an integrity breach. Model extraction and side-channel attacks may reveal sensitive training information or model behavior. Attackers may also inject biased data, modify existing datasets, steal model weights, or reconstruct models using reverse engineering.

 An insider can modify model logic to generate false output. A backdoored model may behave normally until a trigger causes an attacker-favorable response. API-based models also face extraction, inversion, and membership-inference attacks. Third-party xApps/rApps enlarge the software supply-chain risk: a malicious or vulnerable application may elevate privileges, compromise the Near-RT RIC, or launch a fork bomb that continuously replicates and consumes resources until service degrades or fails.

\section{Quantum security solutions for O-RAN}

Quantum computing and quantum information theory use properties of quantum physics to develop advanced solutions for difficult classical problems. There are growing concerns about public-key infrastructures in the face of future quantum attacks against conventional cryptography. At the same time, quantum technologies may provide new protections for network architectures. Possible solutions for O-RAN security include PQC, quantum communication, cryptography, authentication, optimization, and quantum machine learning. Table~\ref{tab:1} shows how these technologies can address specific O-RAN threats. The details are discussed in the following subsections.

\begin{table*}[ht]
 \centering
 \begin{tabular}{|p{3.5cm}|p{3cm}|p{10cm}|}
 \hline
 \textbf{Threat} & \textbf{Target components} & \textbf{ Quantum-safe and Zero Trust defense }\\
 \hline
 Unauthorized access, including spoofing and elevation of privilege attacks & Physical layer, Open fronthaul, O1, E2, A1 interfaces, xApps & Apply continuous verification and least-privilege access; migrate identity and signatures to ML-KEM/ML-DSA. Consider QIA/QDS and QSS/QMC as longer-term complements. \\
 \hline
 Man in the middle attacks at the communication planes & M-plane, U-plane, S-plane, A1 interface & Use quantum-safe mutual authentication, key establishment, and signatures; use QKD-derived keys on selected optical links. \\
 \hline
 PTP attacks & Master clock & Protect timing-source identity and integrity with quantum-safe credentials, redundancy, and monitoring; evaluate QML only as an additional detector. \\
 \hline
 E2 subscription attack & E2 interface & Enforce expiring xApp permissions, subscription quotas, and micro-segmentation; evaluate QML-assisted detection against classical baselines. \\
 \hline
 Denial of Service attacks (Signaling storm, fork bomb attack) & Fronthaul planes, E2 nodes, master clock, core network & Apply admission control, rate limits, workload isolation, continuous monitoring, and policy-based mitigation; QML may provide an additional risk signal. \\
 \hline
 Data poisoning & AI/ML models & Protect data/model provenance with quantum-safe signatures; sanitize and validate training data; evaluate QML-based anomaly detection cautiously. \\
 \hline
 Model attacks (corruption, backdoor attacks) & AI/ML models & Sign model artifacts, enforce least-privilege deployment, and test robust training and rollback; quantum defenses remain experimental. \\
 \hline
 \end{tabular}
 \caption{O-RAN Threats and Quantum Defense.}
 \label{tab:1}
\end{table*}

 \subsection{Zero Trust Architecture for O-RAN}

Zero trust assumes that no user, device, application, vendor, or network location is inherently trusted~\cite{NISTZTA,10495907}. In O-RAN, a policy engine in or adjacent to the SMO can evaluate the identity and security posture of RU/DU/CU functions, O-Cloud workloads, and xApps/rApps. Enforcement points at interface endpoints, API gateways, and RIC platform services then authorize each request. \emph{Continuous verification} re-evaluates long-lived O1, O2, A1, and E2 sessions using certificate status, software measurements, and behavior. \emph{Least privilege} restricts each xApp/rApp to its declared APIs, E2 service models, cells, data, and time window. \emph{Micro-segmentation} isolates RIC applications, O-Cloud tenants, vendors, and management services to limit lateral movement after compromise. Quantum-safe mechanisms strengthen this architecture: PQC protects identities, sessions, software, models, and policies; QKD can supply keys to selected links; and QML can provide an additional risk signal. None of these mechanisms replaces authorization, segmentation, logging, or response.

\subsection{Post-Quantum Cryptography (PQC) for O-RAN}

Before discussing quantum-native hardware, it is essential to address PQC, which provides quantum-resistant security through software-based algorithmic upgrades on classical hardware and existing networks. National Institute of Standards and Technology (NIST) Federal Information Processing Standards (FIPS) 203 standardizes the Module-Lattice-Based Key-Encapsulation Mechanism (ML-KEM) for establishing a shared secret, while FIPS 204 standardizes the Module-Lattice-Based Digital Signature Algorithm (ML-DSA) for digital signatures~\cite{FIPS203,FIPS204}. ML-KEM allows two parties to establish symmetric key material over an untrusted network, whereas ML-DSA enables a sender to sign data so that recipients can verify its origin and integrity; both are based on the hardness of structured lattice problems believed to resist quantum attacks. Within O-RAN, migration should prioritize the management and policy planes. O1 and O2 carry long-lived management data, credentials, configuration, and O-Cloud lifecycle operations, while A1 carries high-level policies; these interfaces are primary targets for ``harvest now, decrypt later'' attacks. Software, firmware, xApp/rApp packages, and AI/ML models also require early quantum-safe signatures because these artifacts may remain deployed for years. E2 needs quantum-safe mutual authentication and xApp handshakes, but Near-RT latency favors using PQC during session setup and efficient symmetric cryptography on the data path. Open fronthaul management and control planes should migrate device authentication and key establishment first; constrained RU hardware, packet size, and synchronization availability require careful testing.

PQC and QKD are complementary. PQC is primarily a software/firmware upgrade and supports both key establishment and signatures over routed interfaces, although its security is computational and its objects are larger. QKD uses a quantum channel plus an authenticated classical channel to distribute symmetric keys and expose eavesdropping attempts. It does not authenticate endpoints by itself, sign software, or protect stored data, and it requires suitable optical links and specialized key management. PQC is therefore the baseline for all O-RAN interfaces, while QKD is a defense-in-depth option for stable, high-value optical paths. O-RAN WG11 has begun work on PQC adoption; concrete algorithm profiles should follow O-RAN/European Telecommunications Standards Institute (ETSI) and 3GPP specifications as they mature~\cite{ORANSec2025}.

Migration requires a cryptographic inventory and algorithm agility. Operators should locate RSA/ECC dependencies in certificates, Transport Layer Security (TLS)/Internet Protocol Security (IPsec)/Secure Shell (SSH), software signing, and enrollment; benchmark ML-KEM/ML-DSA handshake latency, CPU load, message expansion, fragmentation, and hardware-security-module support; and test hybrid classical/PQC handshakes during transition. Downgrade protection, certificate rollover, revocation, and recovery from simultaneous rekeying are as important as single-algorithm benchmarks.

 \subsection{ Mutual Authentication and Authorization}

Open RAN, with its integration of different third-party software, demands strong measures to authenticate parties and validate messages. RAN Intelligent Controllers and xApps/rApps also create permission-management challenges because applications can access and alter network parameters through SMO and RIC interfaces. Two longer-term quantum approaches are quantum identity authentication (QIA) and quantum digital signatures (QDS)~\cite{10574804}. Intuitively, QIA uses quantum states or entanglement-based challenge--response as evidence of identity, whereas QDS uses quantum-generated correlated information to verify message origin and integrity. Both require quantum-capable endpoints or supporting servers and are less mature than standardized PQC. A dedicated Quantum Authentication Server (QAS) could expose authentication evidence to the zero-trust policy engine, while QKD could protect selected communication links.

\begin{figure}[h]
 \centering
 \includegraphics[width=\linewidth]{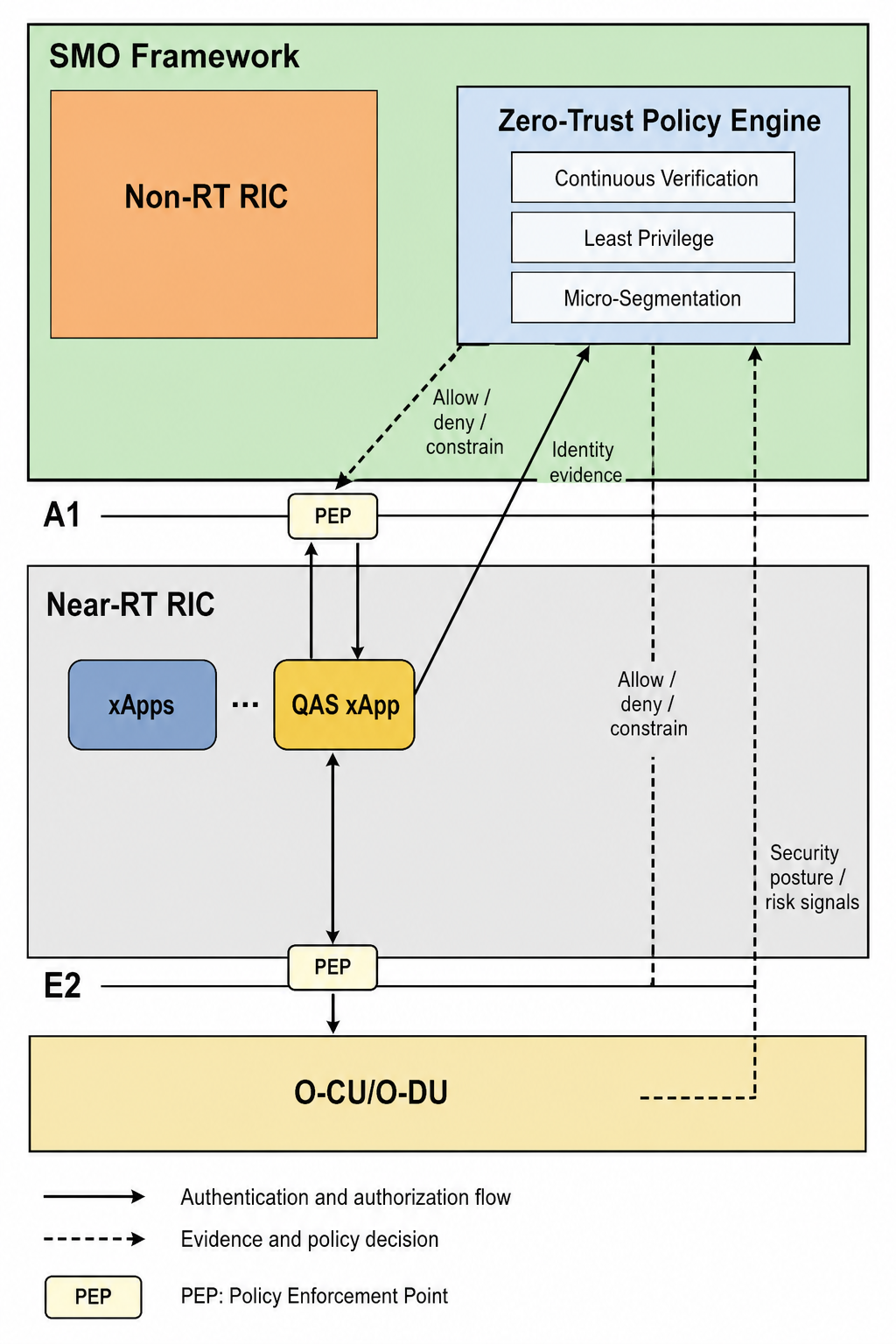}
 \caption{Implementation of Quantum Authentication Server (QAS) as an xApp.}
 \label{fig:1}
\end{figure}

Similarly, a quantum digital signature can provide message authenticity, integrity, and non-repudiation. Unlike standardized PQC signatures, QDS requires quantum-generated correlated information and specialized endpoints. QKD may protect the distribution of symmetric key material, but it does not by itself provide a digital signature or endpoint authentication. A validated quantum random-number generator could improve key generation, although its output and health tests still require secure classical integration.

In \cite{10495907}, the authors proposed to integrate service authentication and secured slicing in xApps to implement stricter control access within the O-RAN. This framework can be leveraged by incorporating QIA to xApps, effectively letting the xApps function as Quantum Authentication Servers. This aligns with the Zero Trust principle of continuous verification: no component remains trusted solely because it passed an initial connection check. This approach protects the network from malicious attackers as both UE and internal network components (other xApps and rApps). Additionally, secured slicing provides continuous authentication and monitoring while limiting the resources allocated to UE to prevent Denial-of-Services attacks.

 When more than two parties collaborate, Quantum Secret Sharing (QSS) acts intuitively like a quantum puzzle: it divides a quantum secret into shares so that only an authorized threshold of parties can reconstruct it, and no single participant or vendor holds the complete secret. Quantum Multiparty Computation (QMC) allows several parties to compute a joint result without revealing their private inputs. In multi-vendor O-RAN, these concepts could protect recovery keys or require joint approval for exceptional configuration changes, supporting least-privilege and separation-of-duty policies. Classical threshold cryptography and PQC can provide similar governance sooner, while QSS/QMC remain candidates for future quantum-network trials.

\subsection{ Privacy Assurance, Confidentiality, and Integrity}

Similar to other quantum cryptography domains, Quantum Key Distribution is a powerful technology for secure communications. QKD protects shared secret keys by leveraging quantum measurement and, in some protocols, entanglement. An eavesdropper's measurement introduces detectable disturbance before the participants accept a key. QKD protocols include prepare-and-measure (e.g., BB84) and entanglement-based approaches (e.g., E91), and may use discrete or continuous variables~\cite{9684555}.

While QKD is a powerful tool for protecting against third-party attacks, thus preventing eavesdropping and man-in-the-middle threats, its implementation in multi-node networks requires extra control and management. In Open RAN architecture, this can be deployed with the help of a QKD control and management layer, or Quantum Key Servers as discussed in \cite{9684555}. From a Zero Trust perspective, assigning independent key domains to network slices or interfaces can support micro-segmentation and limit lateral movement; QKD supplies keys to those domains but does not create segmentation by itself. The control and management layer consists of a QKD controller that directly controls all QKD functions of O-RAN network components and a QKD network manager that supervises all network nodes, including the QKD controller. Quantum Key Servers control and manage subsets of QKD-enabled components based on physical or virtualized infrastructures \and communicate to coordinate key exchange. In both cases, the control units are placed between the physical and application layers to support secure application-layer communication, which is critical to virtualized and cloud-based O-RAN deployments.

\begin{figure}[h]
 \centering
 \includegraphics[width=\linewidth]{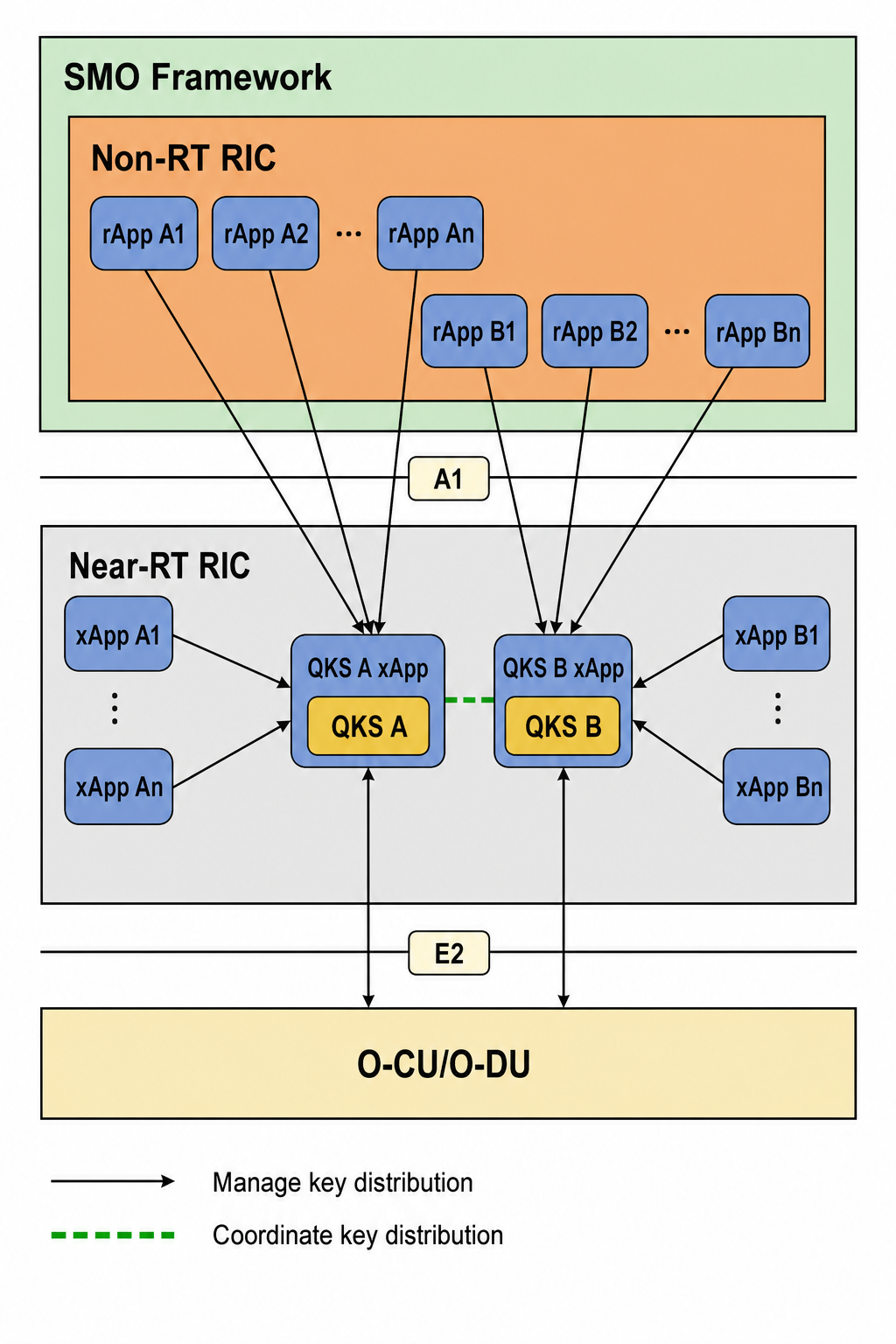}
 \caption{Implementation of Quantum Key Servers (QKSs) as xApps. Each QKS manages a subset of O-RAN components.}
 \label{fig:2}
\end{figure}

\begin{figure}[h]
 \centering
 \includegraphics[width=\linewidth]{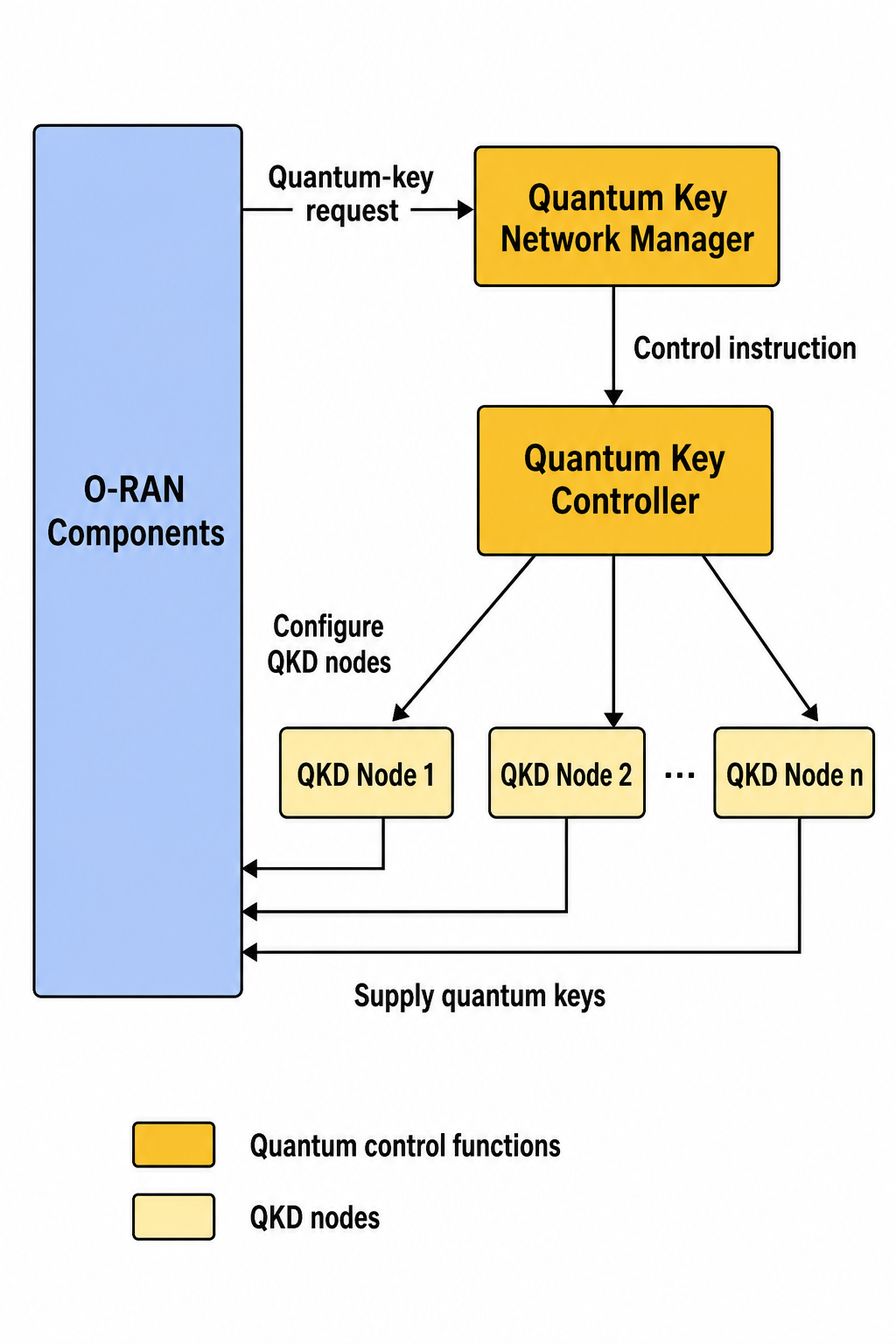}
 \caption{Implementation of QKD as standalone quantum blocks.}
 \label{fig:4}
\end{figure}

Alternatively, Quantum Secure Direct Communication (QSDC) provides secure communication without first distributing a reusable secret key~\cite{9749227}. QSDC encodes a message in quantum states and uses measurement disturbance to detect interference during transmission. Its principal limitations are deployment complexity, distance, and low transfer rate. QSDC may become useful in future quantum networks, but it is not a near-term replacement for PQC or QKD in O-RAN.

O-RAN's multi-vendor architecture requires protection against both external threats and malicious insiders. Quantum protocols may complement Zero Trust by reducing reliance on implicit device or party trust, but they do not replace policy enforcement~\cite{10763508}. Quantum Bit Commitment, where Alice commits to a secret bit value later revealed to Bob, and Quantum Coin Flipping, where Bob adds a bit after Alice's commitment to produce a combined ``coin flip'', are prime examples. While perfectly secure versions of these protocols are proven impossible, weaker variants with reduced cheat probabilities have been proposed. Device-Independent Cryptography offers additional protection for network devices susceptible to exploitation, without relying on assumptions about device trustworthiness.

\subsection{Threat Detection and Mitigation}

Security threats in O-RAN come from various sources, from UE to RAN components and interfaces, as extensively listed in \cite{10467183}. Therefore, O-RAN needs a robust threat and intrusion detection framework capable of near real-time monitoring and intervention while network operations continue. Previously, continuous authentication and monitoring was discussed, which is managed by xApps to prevent malicious UE. However, an intrusion detection xApp was also proposed by \cite{10495907}. With this xApp, network traffic can be monitored continuously to detect and prevent suspicious activities and potential threats. This monitoring implements the continuous-verification pillar of Zero Trust. Quantum technology may further assist high-dimensional analysis, although any advantage must be demonstrated end to end against classical approaches.

Quantum Machine Learning (QML) integrates quantum mechanics with machine learning to perform supervised, unsupervised, and reinforcement learning, leveraging expanded computational bases and parallel computation to improve upon classical methods \cite{10233127}. Most QML algorithms require encoding classical data into quantum bits before applying quantum circuit gates. For intrusion detection, quantum-based support vector machines, decision trees, and random forests can identify suspicious activity from known attack patterns, with xApp performance measurements encoded as qubits for quantum circuit processing. For anomaly detection, quantum Gaussian mixture models represent cluster distributions as wave functions, enabling parallel quantum computation to efficiently identify outliers. Quantum reinforcement learning further supports continuous parameter training through feedback loops, making it well-suited for the O-RAN AI/ML framework where network parameters stream constantly through the O1, A1, and E2 interfaces as xApp training data.
\begin{figure}[h]
 \centering
 \includegraphics[width=\linewidth]{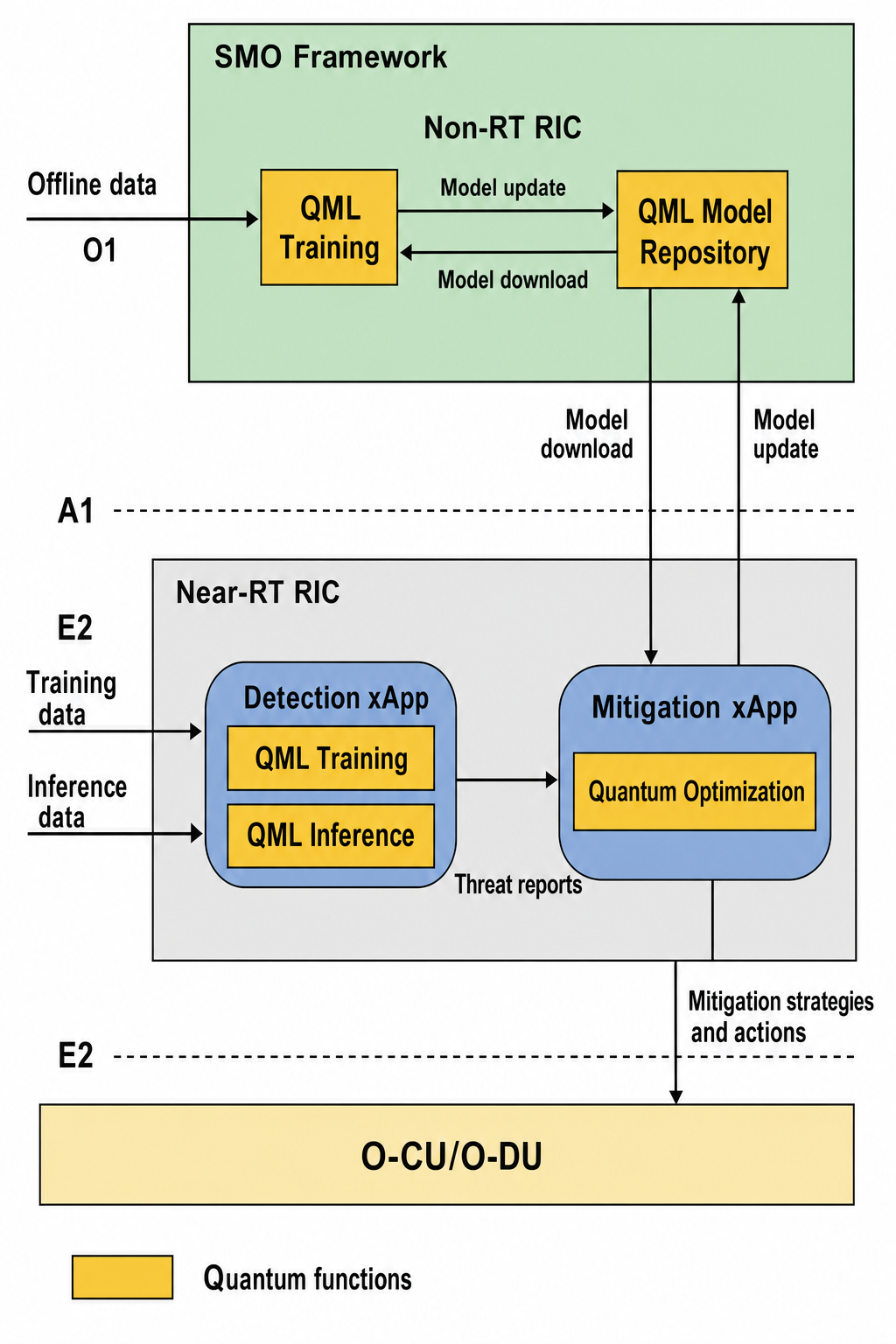}
 \caption{Implementation of Quantum Machine Learning (QML) for threat detection/mitigation as xApps.}
 \label{fig:3}
\end{figure}

Additionally, various defense mechanisms rely on optimization problems to minimize exposure and risks while retaining network availability and functionality. Hence, quantum optimization is a highly promising tool for this task. At its core, quantum optimization takes advantage of unique resources and strategies for modeling and solving conventional optimization problems to fully realize the power of quantum machines. For O-RAN, a problem like signaling storm could be mitigated by blocking certain UEs from subscribing to the network. This problem can be framed as optimizing the number of malicious UEs blocked out of all users while also keeping check of the number of legitimate UEs wrongly blocked.

\subsection{ Poisoning Attacks}

Poisoning attacks target the training phase of the model to stealthily corrupt its parameters from within. Although the damage is not immediately obvious, contrary to inference attacks, poisoning attacks have significantly more widespread and long-lasting impacts on the system. Since the target is training data, the best solution is to implement multi-layered protections for the dataset and training process. However, due to the continuous flow of real-time data in O-RAN, data protection and sanitization is difficult. With the help of anomaly and threat detection xApps in the network as discussed above, abnormal data points should be appropriately identified and removed from the training set. Ideally, these xApps should rely on quantum clustering algorithms like quantum Gaussian mixture and k-nearest neighbours that require minimal pre-training to avoid data poisoning. Quantum cryptography techniques such as QKD and QSDC can be used to protect the live data transmission via the O1, A1 and E2 interfaces, preventing attackers from obtaining and modifying the data feed to the RICs. Furthermore, an xApp for integrated risk assessment and threat detection can monitor other xApps on the same network to detect any misbehaviours, either from poisoning attacks or already corrupted models. Given the complexity of these tasks, Quantum Reinforcement Learning should be considered to enhance and adapt the model to new and potential threats. Possible implementations include quantum-based Q-learning, experience replay, and actor--critic methods. In an actor--critic architecture, the actor selects actions and the critic evaluates them; quantum variants remain experimental and should not be assumed to provide an exponential advantage without end-to-end evidence.

For AI/ML models with offline training (models in the Non-RT RIC or anomaly detection xApps trained on existing data), efforts should be dedicated to upholding the dataset's integrity and privacy. Conventional approaches involve blockchain, post-quantum cryptography, data masking, and sanitization \cite{11018396}. Consequently, effective quantum technologies include QDS for authentication and auditing, QKD for security, and QSS/QMC for collaborative authorization. Many data-driven countermeasures formulate the removal of poisoning samples as an optimization problem. Quantum search or optimization may eventually assist particular formulations, but Shor's factorization algorithm is not a poisoning defense and no generic operational speedup has been established. These methods also risk misidentifying legitimate data, increasing resource consumption, and reducing accuracy on complex datasets. Model-driven approaches instead improve robustness within the learning scheme. Nguyen et al.~\cite{11018396} discuss quantum adversarial training, quantum defense-generative adversarial networks (GANs), defensive distillation, and gradient masking as candidate countermeasures. These methods remain experimental and require evaluation against adaptive attacks and strong classical baselines.

\section{Integration of quantum nodes in O-RAN}
Quantum features such as quantum cryptography require advanced hardware to function properly. Even today's most cutting-edge quantum chips are inadequate to fully realize the power of quantum technologies. However, transitional techniques like local operations and classical communication (LOCC) can bridge the gap between traditional and quantum information practice, allowing researchers to propose new quantum nodes in future network architectures. With this in mind, the previously discussed quantum nodes including quantum key controller, quantum key manager, quantum key server and QML modules can be implemented into the O-RAN as demonstrated by table \ref{tab:2}.

\begin{table*}[ht]
 \centering
 \begin{tabular}{|p{3cm}|p{5cm}| p{2cm}|p{6cm}|}
 \hline
 \textbf{Quantum node} & \textbf{Function} & \textbf{Component} & \textbf{ O-RAN control path}\\
 \hline
 Quantum authentication server & Process authentication request to verify legitimate users and prevent unauthorized access to network. & Standalone/xApp & Receive authentication request from O-RAN components through open interfaces (xApp) or dedicated interfaces (standalone) and send back the results. \\
 \hline
 Quantum key network manager & Manage key requests and direct them to the controller. & Standalone/xApp & Process secret-key request from O-RAN components, perform key query and verification before forwarding them to the controller. \\
 \hline
 Quantum key controller & Control and manage quantum keys and QKD between O-RAN components and QKD nodes. & Standalone/xApp & Receive instructions from the key manager and configure the corresponding QKD nodes to deliver secret keys to the target O-RAN component through separate interfaces. \\
 \hline
 QKD node & Generate quantum secret key and supply them to target O-RAN component as instructed by the controller. & Standalone & Receive request from controller, check availability for real-time quantum key and send the keys to designated O-RAN components through dedicated interfaces. \\
 \hline
 Quantum key server & Supervise quantum key distributions of an assigned network slice. & Standalone/xApp & Coordinate key exchange request with assigned network components via its specific interface (E2, O1 and A1 for xApp) to establish key distribution between nodes through open interfaces. \\
 \hline
 Quantum threat and intrusion detection & Identify and report suspicious activities from live network performance measurements using QML models. & xApp & Receive live performance measurement data via the E2 interface to the near-RT RIC. Infer suspicious activities and anomalies from the data and report threats and intrusion to the mitigation module. \\
 \hline
 Quantum threat mitigation module & Deploy optimized defense strategies against network attacks reported by the detection module. & xApp & Receive threat reports from the detection xApp and initiate appropriate responses through the near-RT RIC block. \\
 \hline
 \end{tabular}
 \caption{Quantum nodes implementation in O-RAN.}
 \label{tab:2}
\end{table*}

\subsection{Quantum Nodes as xApps}

Many quantum features can be fully integrated into xApps as network services in the O-RAN, allowing for third-party software. Since xApps can subscribe to decentralized RAN units to access live data and make adjustment, they are a powerful tool for network-wide monitoring and control. For example, the quantum authentication and key distribution schemes can be integrated into O-RAN following Figs.~\ref{fig:1} and \ref{fig:2}. Similarly, quantum machine learning models can reside within xApps as demonstrated by Fig.~\ref{fig:3}. By hosting quantum features inside xApps, third-party vendors can ensure their software is compatible with the O-RAN architecture for easy maintenance and testing.

\subsection{Dedicated Quantum Blocks in O-RAN}

While implementing quantum nodes inside xApps allows for ease of deployment and integration with the O-RAN, it also raises security concerns in cases where network interfaces and components are compromised. Furthermore, this strategy limits third-party software to predefined interfaces. Thus, another less restrictive approach is to add standalone components to the O-RAN using new nodes and interfaces. This alternative implementation of quantum key distribution can be adopted following Fig.~\ref{fig:4}.

Here, the QKD nodes are responsible for generating and storing the secret keys of each individual O-RAN component, allowing third-party QKD software and strategies independent of the main network. However, extra efforts must be dedicated to test and monitor these external components to ensure the core network is safe from both malicious insiders and outsiders. Finally, setting up components outside of the predefined O-RAN architecture may cause conflicts and maintenance issues if not regulated properly.

\subsection{Maturity and Implementation Roadmap}

Adoption should proceed in phases based on technological maturity. \textbf{Short term (PQC transition, approximately 1--3 years):} operators can inventory RSA/ECC use, introduce crypto-agile APIs, test hybrid classical/PQC handshakes, migrate software and model signing, and enforce Zero Trust controls. ML-KEM and ML-DSA run on classical O-Cloud and RIC infrastructure, but their larger keys, ciphertexts, and signatures must be benchmarked on RU, DU, RIC, SMO, and O-Cloud platforms. Mixed-vendor testing should cover certificate enrollment, rollover, revocation, downgrade resistance, and recovery after a rekeying storm.

\textbf{Medium term (hybrid PQC--QKD):} as O-RAN, ETSI, and 3GPP profiles and product support stabilize, PQC can be deployed broadly while QKD is introduced on selected high-value optical links, including appropriate SMO--RIC or transport paths. Pilots must measure secret-key rate, loss budget, distance, key-consumption rate, failover time, availability during fiber maintenance, trusted-node exposure, and tenant isolation. If a quantum link fails, policy-controlled fallback to PQC-established keys is necessary to avoid a service outage.

\textbf{Long term (native quantum capabilities, potentially 7+ years):} broader QIA/QDS, QSS/QMC, QSDC, quantum networking, and production QML depend on scalable quantum hardware, repeaters, and standardized quantum--classical interfaces. QML trials should operate beside classical detectors and compare end-to-end detection accuracy, false-positive rate, latency, data-loading overhead, energy, and robustness to poisoning~\cite{11018396}. Quantum controllers, key managers, and authentication servers must themselves be redundant, attested, least-privileged, monitored, and protected against denial-of-service attacks. These ranges are planning horizons, not predictions of when a cryptographically relevant quantum computer will exist.

 \section{Conclusion}
The rapid evolution of O-RANs toward disaggregated, multi-vendor architectures introduces significant security challenges that conventional cryptographic mechanisms alone cannot adequately address, particularly in the face of emerging quantum-enabled adversaries. By combining a Zero Trust framework with quantum-resilient tools, this article has presented a framework for integrating quantum-based security solutions into O-RANs, encompassing Post-Quantum Cryptography (PQC), quantum identity authentication, quantum key distribution, and quantum machine learning. By mapping specific O-RAN threats to corresponding quantum defense mechanisms, and proposing practical deployment models via intelligent controller applications and standalone quantum nodes, this work bridges the gap between quantum information science and open wireless network architectures. While current quantum hardware limitations remain a barrier to full deployment, transitional approaches such as the PQC-first roadmap offer viable near-term pathways. As quantum technologies continue to mature alongside sixth-generation standardization efforts, the integration of quantum-resilient mechanisms into O-RANs will be essential for building secure, trustworthy, and future-proof next-generation wireless infrastructure.

\bibliographystyle{IEEEtran}
\bibliography{bibiliography_revised}
\begin{IEEEbiographynophoto}{D. Q. Ngo} received his MSc in Artificial Intelligence from Sheffield Hallam University in 2024. He is currently a Research Assistant at London Digital Twin Research Centre, Middlesex University, London, U.K. His research interests include LLMs, multimodal AI, and neuro-symbolic AI.
\end{IEEEbiographynophoto}
\vskip -2\baselineskip plus -1fil
\begin{IEEEbiographynophoto}{T. Raveendranathan} received her B.S. in Electrical and Electronic engineering from University of Peradeniya, Sri Lanka in 2019, PGDip in Electrical Installations from University of Moratuwa, Sri Lanka, in 2023 and MSc. in Telecommunication Engineering from Middlesex University in 2024. She worked as a research assistant at London Digital Twin Research Centre. Her research interests are digital twin, bio-inspired optimization algorithms, RIS, and O-RAN security.
\end{IEEEbiographynophoto}
\vskip -2\baselineskip plus -1fil
\begin{IEEEbiographynophoto}{T. A. Le} received his Ph.D. in Telecommunications from King's College London in 2012 and was a Post-Doctoral Fellow at the University of Leeds. He is now a Senior Lecturer at Middlesex University. His research spans integrated sensing and communication, energy harvesting, physical-layer security, and ML for wireless communications. Le was honored as an Exemplary Reviewer by IEEE Communications Letters in 2019/2023 and serves as an Editor of the IEEE Wireless Communications Letters. He received the Best Editor Award of IEEE Wireless Communications Letters in 2025.
\end{IEEEbiographynophoto}
\vskip -2\baselineskip plus -1fil
\begin{IEEEbiographynophoto}{V. Sharma} received his BTech in EE from IIT Delhi in 1978 and PhD in ECE in 1984 from Carnegie Mellon Univ at Pittsburgh, USA. After Ph.D., he taught in Northeastern University and University of California at Los Angeles before joining Indian Institute of Science (IISc) in Bangalore. He has been Dept Chair of ECE and Dean of Engineering in IISc before super-annuation in 2022. Since then he is a Prof of EE in Shiv Nadar Institution of Eminence in Greater Noida, UP, India. His research interests include Comm Networks, Information Theory and Quantum Technologies.

\end{IEEEbiographynophoto}
\vskip -2\baselineskip plus -1fil
\begin{IEEEbiographynophoto}{P. Shah} is the Director of Programmes for Engineering programmes and a Senior Lecturer at Middlesex University London, where he also leads the MSc Robotics programme. He received his Ph.D. in Computing and Electronics Engineering from the University of Plymouth in 2008. His research interests include IoT system modelling, digital twins, autonomous networks, intelligent transportation systems (ITS), AI-enabled smart systems, and next-generation wireless communications and sensing. He has published over 85 research articles with collaborators in UK, India and Europe; and has secured UK research funding in Industry 4.0, digital twins, and ITS, and serves on IEEE conference and journal committees, as well as the UKRI EPSRC, MRC, and Innovate UK peer review college.
\end{IEEEbiographynophoto}
\vskip -2\baselineskip plus -1fil
\begin{IEEEbiographynophoto}{H. X. Nguyen}
is a Professor of Digital Communication Engineering, Director of the London Digital Twin Research Centre and Head of the 5G \& IoT Research Group at Middlesex Univ., London, UK. He is also a Visiting professor at VinUniversity, Hanoi, Vietnam. He received his Ph.D. degree from the Uni. of New South Wales, Australia, in 2007. He leads research activities in digital twinning, 5G/6G/IoT systems, and digital transformation solutions.
\end{IEEEbiographynophoto}
\vskip -2\baselineskip plus -1fil
\end{document}